\documentclass[aip,reprint,superscriptaddress,eqseqnum,author-year]{revtex4-1}
\usepackage{amsmath,amsthm,amsfonts,amssymb}
\usepackage{graphicx}
\usepackage{color}
\usepackage{natbib}
\usepackage{mciteplus}

\renewcommand{\vec}[1]{{\mathbf #1}}        
\newcommand{\uvec}[1]{\widehat{\mathbf #1}}
\newcommand{\ten}[1]{{\mathbf{#1}}}
\newcommand{\half}{\textstyle{\frac{1}{2}}}

\newcommand{\Id}{\ten{I}}
\newcommand{\thetabold}{\mbox{\boldmath{$\theta$}}}
\newcommand{\note}[1]{{[\small\textbf{\texttt{\color{red}#1}}]}}

\newcommand{\change}[1]{{\color{black}{#1}}}   

\begin{document}
 
\title{Transient shear banding in entangled polymers: a study using the Rolie-Poly model}

\author{J.~M.~Adams} \affiliation{SEPNet and the Department of Physics, University of
  Surrey, Guildford, GU2 7HX, UK}

\author{S. M. Fielding}
\affiliation{Department of Physics, Durham University, Science
  Laboratories, South Road, Durham, DH1 3LE, UK}

\author{P.~D.~Olmsted}
\affiliation{ School of Physics \& Astronomy, University of Leeds,
  Leeds, LS2 9JT, UK} 
\date{\today}
\begin{abstract} {Spatially inhomogeneous shear flow occurs in
    entangled polymer solutions, both as steady state shear banding
    and transiently after a large step strain or during start up to a
    steady uniform shear rate.  Theoretically, steady state shear banding is a
    hallmark of models with a non-monotonic constitutive relation
    between total shear stress and applied shear rate, but transient
    banding is sometimes seen in fluids that do \textit{not} shear
    band at steady state. We model this behavior using the diffusive
    Rolie-Poly model in a Newtonian solvent, whose steady state constitutive
    behavior can be monotonic or non-monotonic depending on the degree
    of convective constraint release (CCR). We study monotonic
    steady state constitutive behaviour.  Linear stability analysis of start up to
    a sufficiently high shear rate shows that spatial fluctuations are
    unstable at early times.  There is a strong correlation between
    this instability and the negative slope of the (time dependent)
    constitutive curve. If the time integral of the most unstable
    eigenvalue is sufficiently large then the system exhibits
    transient shear bands that later vanish in steady state. We show
    how perturbations, due to fluctuations or the inhomogeneous
    stresses, can trigger this instability. This transient behavior is
    similar to recent observations in entangled polymer solutions. }
\end{abstract}
\maketitle
\section{Introduction}

The original theory of \citet{doiedwards} predicts shear banding in
entangled polymers, when the shear rate exceeds the reciprocal of the timescale $\tau_d$
for one-dimensional diffusion (reptation) within an effective mean
field tube of constraints from the surrounding molecules. For shear
rates $\dot{\gamma}\agt1/\tau_d$, DE theory predicts a decreasing
shear stress $\sigma_{xy}(\dot{\gamma})$ due to enhanced tube
alignment, and thus leads to a non-monotonic constitutive curve
$\sigma_{xy}(\dot{\gamma})$ and an instability to an inhomogeneous
state in which two states, or `shear bands', of different viscosity
and molecular alignment coexist \citep{mcleish86}.  However, banding
was not inferred at that time in polymers; for example, the apparent
viscosity measurements of \citet{menezes1982nonlinear} on entangled
polymers are consistent with a weakly increasing shear stress for
$\dot{\gamma}\tau_d>1$, rather than the stress plateau that is
expected from banding. By contrast, banding is prevalent in wormlike
micelles
\citep{rehage91,berretrev05,catesfielding06,olmstedbanding08}. This
was explained by \citet{cates90}, who combined reptation with micellar
breaking to predict shear banding at stresses and shear rates in good
qualitative agreement with experiments \citep{spenley93}. It is now
realized that a strictly constant stress plateau is not obtained for
banding in a curved geometry, such as cone and plate or cylindrical
Couette fixtures \citep{olmsted99a}, so the results of
\citet{menezes1982nonlinear} may in fact be consistent with shear
banding.

Discrepancies with DE theory have existed virtually since the theory's
inception \citep{fukuda1975nonlinear}, particularly in the response of
highly entangled polymeric fluids to a step
strain. \citet{osaki1980experimental} and \citet{vrentas1982study}
observed stress relaxation that was much faster than predicted by DE
theory. This is particularly prevalent for entanglement numbers
$Z\agt60$ for relaxation after large step strain
\citep{osaki1993damping}, and has been termed anomalous or ``type C''
relaxation \citep{osaki1993damping,venerus2005critical}.
\citet{marrucci1983free} (MG) pointed out that the DE theory contains
an elastic instability; they showed that the free energy can be a
convex function of strain for large enough step strain (for
$\dot{\gamma}\tau_d\gg1$). This would then lead to an elastic
instability and thus inhomogeneous (shear banding) behavior. They
showed how this instability can persist even when one includes
relaxation.
   \citet{mcleish86} later suggested
that the DE theory could explain the measurements of the spurt effect
by \citet{vinogradov73}, in terms of the inherent instability to shear
banding. \cite{kolkka1988spurt} used the Johnson-Segalman model to
study the nature of mechanical instabilities inherent in fluid with a
non-monotonic constitutive relation. It was later demonstrated that
wall slip plays a major role in spurt \citep{Wan1999PCE,Den2001ARFM}.
\citet{MorrLars92} tested Marrucci and Grizzuti's idea by using
entangled solutions at different concentrations to mimic the effect of
relaxation. They concluded that the Marrucci-Greco mechanism gives a qualitatively
correct description of the anomalous relaxation, which is consistent
with incipient inhomogeneities. More recently,
Venerus \citep{venerus2005critical,venerus2006stress} studied step
strain experiments, and concluded that many of the anomalous ``type
C'' responses could be explained by experimental conditions such as
wall slip or rheometer compliance; however, a number of experiments
could not be accounted for in this way.

In the last decade or so the advent of high resolution velocimetry has
led to a series of exciting experiments, principally by Wang and
co-workers, that demonstrate evidence of shear banding in entangled
polymers
\citep{tapadiawang03,Tapadia2004Nonlinear-flow-,boukany07,hu08,ravindranath2008banding}.
Very well-entangled solutions, with entanglement number $Z\agt40-50$,
display well-defined steady state shear bands
\citep{wang2006nonquies,hu2007cre,hu08,ravindranath2008banding,boukany2009shear}. This
occurs in both synthetic polymers, such as polybutadiene
\citep{ravindranath2008banding}, and DNA
\citep{WangMM2008,boukany2009exploring,hu2008role,boukany2009shear}. Polydisperse systems exhibit
smoother, less well-defined (or no) shear banding \citep{boukany07},
possibly because the many timescales smear out the instability (as
suggested by \citet{doi1979dcp} in their original paper).


There have been numerous improvements upon the original DE theory,
notably to incorporate convective constraint release (CCR)
\citep{marrucci96,mead98,MilnerML01}. CCR can reduce the severity of
the DE instability and potentially render the fluid stable
\citep{graham2003microscopic}, and DE-CCR models can capture, at least
qualitatively, many of the shear banding signatures of these
experiments. This was recently shown by
\citet{adamsolmsted09a,adamsolmsted09b} in calculations
based on the Rolie-Poly model \citep{likhtmangraham03}, a simple
one-mode differential version of DE theory that incorporates CCR and
chain stretch.  However, there is no consensus yet as to the correct
level of CCR required to describe existing experiments.  

In related work, \citet{cook08} showed that a non-monotonic
constitutive relation based on the ``partially extending convective
strain'' model of \citet{larson1984constitutive} also reproduces many
of the features of the recent experiments, including recoil during
startup onto the banding plateau, and strain localization immediately
after a step strain effected by shearing at a very fast finite shear
rate.

Numerous experiments also show inhomogeneous behavior in fluids that
do not display steady state shear banding  (smaller
entanglement number $Z$). Examples include: (1)
transient band formation and recoil during startup, which eventually
gives way to homogeneous shear flow
\citep{tapadiawang06,hu2007cre,ravindranath2008banding,boukany2009shear};
(2) extremely sharp shear banding during portions of the cycle during
Large Amplitude Oscillatory Shear (LAOS)
\citep{tapadiaravin06,ravindranath2008large,zhouLAOS2010}; (3) inhomogeneous
response (including negative velocity recoil) after  step
strain performed at high shear rates \citep{wang2006nonquies,WangMM2007}.

At a constitutive level, one cause of steady state shear banding is a
non-monotonic steady state constitutive relation (such as the DE
theory with sufficiently weak CCR), which cannot support stable
homogeneous steady states for a range of shear rates. However, the
same criterion clearly cannot determine the apparent instability to
transient banding in fluids that don't shear band in steady state
(\textit{i.e.} that have monotonic flow curves). Evidently these
fluids are dynamically unstable, which need not be related to a steady
state non-monotonic constitutive relation. 

\change{Indeed, pronounced transient shear banding has recently been reported
by \citet{DivTam2010PL} in a yield stress fluid with a monotonic steady state constitutive
curve, and captured theoretically in shear transformation
zone (STZ) theories \citep{ManLan2007PSNSMP,manning2009rate}, a modified soft glassy rheology
(SGR) model \citep{moorcroft2011age}, a simplified fluidity model \citep{moorcroft2011age}, and a
mesoscopic model of plasticity \citep{jagla2010towards}.}

Several, possibly related,
explanations for the transient banding can be envisioned:
\begin{enumerate}
\item The dynamical equations of motion passes through a regime of
  parameter space, as a function of time, in which the homogeneous
  instantaneous state is dynamically unstable to small spatial
  perturbations. Such perturbations will initially grow in time before
  eventually decaying to a homogeneous steady state. However, if the
  instability is strong enough the perturbations could grow into a
  macroscopically observable transient band before decaying. We will
  perform this linear stability calculation here, and show that this
  gives an understanding of the transient homogeneities. A similar
  calculation was done by \citet{fielding03c} for wormlike micelles
  undergoing steady state banding.

\item The fluid possesses, at any observation time $t_m$, an
  \textit{instantaneous} constitutive curve
  $\sigma_{xy}(\dot{\gamma},t_m)$. Such a constitutive curve can be
  constructed by performing a number of startup evolutions
  (calculations or experiments) at different shear rates, which then
  define the loci of stress as a function of shear rate for given
  observation times $t_m$ after flow inception. Since most of these
  points are not in steady state, the resulting curve need not be
  monotonic even if the shear rate remains homogeneous.  Then, one
  could associate a negative slope $\partial
  \sigma_{xy}(\dot{\gamma},t_m)/\partial\dot{\gamma}<0$ with transient
  inhomogeneities. However, this only addresses a subspace of the full
  parameter space considered by the first scenario above and could
  thus only serve as a rough guideline. One must also carefully
  distinguish between theoretical calculations in which a homogeneous
  shear rate can be specified, and the experimental protocol above, in
  which the fluid will be expected to become inhomogeneous after the
  onset of an instability. This procedure was recently suggested by
  \citet{hayes2010constitutive} in experiments using parallel plate
  rheometry of polymer solutions.

\item For very strong shear rates these viscoelastic fluids respond
  elastically, like a non-linear solid, and can display a stress
  overshoot. In this case the mechanism of \citet{marrucci1983free}
  {might be expected to apply, as follows.}  For a given imposed shear
  rate $\dot{\gamma}$ the stress $\sigma_{xy}$ is a function of
  \textit{strain} $\gamma=\dot{\gamma}t$. A solid with a negative
  gradient $\partial \sigma_{xy}/\partial\gamma<0$ has an effective
  negative differential shear modulus, which leads to elastic
  instability. This would imply that, for shear rates large enough to
  remain in the elastic regime, a stress overshoot upon startup could
  signify an instability to inhomogeneous flows. {[Note that the MG
    argument is strictly for an ideal step strain, rather than a
    strain incurred during a fast but finite shear rate].}
  \citet{sui2007iep} pointed this out in their recent visualization
  study of the same materials studied by \cite{tapadiawang06}.  This
  scenario is suggested by data showing correlations between transient
  banding and stress overshoots (\textit{e.g.} the Figures~3 and~4 of
  \citet{WangMM2008}), and by calculations that show similar
  correlations (\textit{e.g.} Figure 3 of \citet{adamsolmsted09a}, for
  both monotonic and non-monotonic constitutive curves).
 \end{enumerate}
 
 In this paper we explore these different scenarios, and thus the
 detailed conditions necessary for transient inhomogeneities, using
 the diffusive Rolie-Poly (DRP, or RP) model as an example
 \citep{likhtmangraham03}. We calculate its instantaneous linear
 stability for parameters ranging from monotonic to non-monotonic
 constitutive behavior. We show that even fluids with monotonic
 constitutive curves, which have stable homogeneous steady states, can
 have periods of instability during startup flow. For some parameters,
 strong linear instability is shown to lead to transient banding and
 negative velocity recoil in the full non-linear dynamics.

 Since instabilities require an initial perturbation to manifest in
 eventual non-linear growth, we will study two natural perturbations:
 (1) stress gradients, as found in typical circular rheometric devices
 (cone and plate or cylindrical Couette), and (2) non-uniform (noisy)
 initial conditions. We find a strong link between transient
 inhomogeneities, stress overshoots, and an instantaneous
 non-monotonic constitutive curve, but we leave further detailed
 analysis of this for future work.

 We compare our calculations with recent experiments on transient data
 after startup to steady state. For conciseness, we do not study the
 inhomogeneous response to a step strain, though our results give a
 quantitative method for understanding this behavior as well.

\section{The Diffusive Rolie-Poly Model}
\subsection{Momentum Balance}
Newton's force balance for the fluid is given by
\begin{equation}
  \rho \frac{D \vec{v}}{Dt} = \nabla\cdot \boldsymbol{\sigma}
\label{eqn:divstress}
\end{equation}
where $\rho$ is the density, $\boldsymbol{\sigma}$ is the total
stress, and $D/Dt$ is the material derivative. We will use the full
equation set {(at finite but small Reynolds number)} for the linear
stability analysis in section \ref{sec:LSA}, but take the zero
Reynolds number (creeping flow) limit for calculating the full
spatially resolved non-linear dynamics in section \ref{sec:SR}. In the
latter case the equation of motion reduces to
\begin{equation}
\nabla\cdot \boldsymbol{\sigma} = 0.
\label{eqn:creep}
\end{equation}
For the experimental cases of interest we expect very small Reynolds number
$\textrm{Re}\ll10^{-3}$.
\subsection{Stress}
The total stress $\boldsymbol{\sigma}$ in a polymer solution is
assumed to comprise an elastic stress carried by the backbone of the
polymers (due to their stretching and orientation), and viscous drag
against solvent and other polymers. In polymer melts the viscous
stress can usually be safely neglected, particularly for weak flows
\citep{doiedwards}. However, for strong flows, where DE theory
predicts a strongly decreasing polymeric stress due to tube alignment,
it is necessary to incorporate additional viscous stresses from the
faster degrees of freedom. This includes the Newtonian solvent
viscosity of solutions, as well as fast Rouse modes and interpolymer
viscous friction neglected in the simplest tube models. In addition to
this crucial argument, a second contribution to the total stress is
physically necessary to describe steady state shear banding in planar
Couette flow, in which the shear rate is inhomogeneous but the total
stress must be homogeneous.

Hence, we represent the total stress as two separate components: fast
Newtonian (or solvent) degrees of freedom, and the slow viscoelastic
stress $G \ten{W}$:
\begin{equation}
\boldsymbol{\sigma} = -p \Id + 2 \eta \ten{D} + G \ten{W},
\label{eqn:totalstress}
\end{equation}
where $\Id$ is the identity tensor, $\ten{D} = \half\left[\nabla
  \vec{v} + (\nabla \vec{v})^T\right]$, $\vec{v}$ is the velocity
field, $p$ is the isotropic pressure determined by incompressibility
($\nabla \cdot \vec{v} = 0$), $\eta$ is the solvent viscosity, and $G$
is the plateau modulus. In this representation the quantity $\ten{W}$
is the polymer, or viscoelastic, strain, whose stress $G\ten{W}$ is
parametrized by the elastic modulus $G$. Together with a DE-like
constitutive relation for $\ten{W}$, Eq.~(\ref{eqn:totalstress}) can
yield either a monotonic or a non-monotonic constitutive relation,
with a Newtonian high shear rate branch.
\subsection{Rolie-Poly Model}
There are many possible constitutive models for the polymeric strain
$\ten{W}$ \citep{islam2001nonlinear,mead98}, but here we will use the
diffusive Rolie-Poly (DRP) model of \citet{likhtmangraham03}. This is
a single mode approximation to the GLAMM model
\citep{likhtmanMM00,MilnerML01,graham2003microscopic}, and includes
convective constraint release (CCR), reptation of the polymers within
their tubes, and the stretching of the polymer chains. The
constitutive equation for the deviatoric part of the viscoelastic
strain $\ten{W}$ is
\begin{multline}
  (\partial_t + \vec{v}\cdot \nabla) \ten{W}-\left(\ten{\nabla
      v}\right)^T \cdot \ten{W} - \ten{W}\cdot (\ten{\nabla
    v}) + \frac{1}{\tau_d}\ten{W} =\\ \label{eqn:RPmodel} 2
  \ten{D} - \frac{2}{\tau_R} \left( 1-A\right) \left( \Id
    +\ten{W} + \beta A^{-2\delta} \ten{W}\right) + \mathcal{D} \nabla^2
  \ten{W},
\end{multline}
where
\begin{equation}
A(\ten{W}) = (1+\mathrm{ tr} \ten{W}/3)^{-1/2},
\end{equation}
$\tau_d$ is the disengagement time, $\tau_R$ is the Rouse time, and
$\beta$ parametrizes the efficiency and rate of CCR.  The ratio of
disengagement to Rouse times defines the entanglement number
\citep{larson2003definitions}
 \begin{equation}
\label{eq:entnumber}
 Z = \frac13\frac{\tau_d}{\tau_R}.
 \end{equation} 
 We have added `diffusion' (the term with coefficient $\mathcal{D}$)
 to the original Rolie-Poly model, to be able to resolve spatial
 structure during shear banding \citep{lu99,olmsted99a}. The width of the interface between shear bands is proportional to $\sqrt{\mathcal{D}}$.
\begin{figure}
\includegraphics[width =0.48\textwidth]{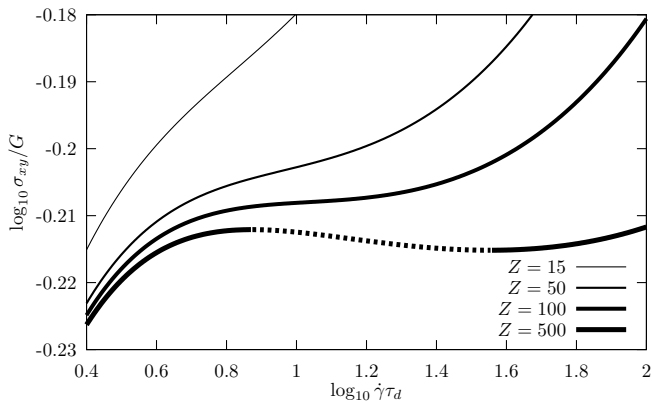}
\caption{Constitutive curves for the Rolie-Poly model, showing monotonic and 
non-monotonic behavior as a function of  entanglement number $Z$;
for $\beta=0.65$ and $\epsilon=10^{-5}$.}
\label{fig:CCZs}
\end{figure}
 
The RP model does not reproduce all the same constitutive behavior
features of the full GLAMM constitutive model (notably, the high shear
rate behaviors of the normal and shear stresses differ), primarily
because only a single mode of the GLAMM model has been kept in
obtaining the RP model. However, it is simple enough to perform
spatially resolved simulations in the presence of the added diffusive
term. The additional solvent stress, $2\eta\ten{D}$ in
Eq.~(\ref{eqn:totalstress}), was not included in the original
formulation, which was focused on melts and not on shear banding
behavior.

\subsection{Choice of Parameters}
\citet{likhtmangraham03} chose $\beta=0.5$ and $\delta=-1/2$, which
optimized the comparison with transient and steady state measurements
within the assumption of homogeneous flows. The negative value for
$\delta$ ensures that CCR decreases for large stretch, due to the
increased number of effective entanglements.  The high shear rate
scaling of the total shear stress $\sigma_{xy}$ in the RP model is
\begin{equation}
\frac{\sigma_{xy}}{G}
= \begin{cases}\left[\frac{\displaystyle1}{\displaystyle(6Z+1)^2} +
    \epsilon\right]\dot{\gamma}\tau_d & (\delta\leq0)\\[12truept] 
\left(\frac{\displaystyle3^\delta \dot{\gamma}\tau_d}{\displaystyle 6
    Z \beta 2^{\delta}}\right)^{\frac{1}{1+2\delta}} +
\epsilon\dot{\gamma}\tau_d& (\delta > 0), 
\end{cases}
\label{eqn:Tlimits}
\end{equation}
where 
\begin{equation}
\epsilon=\frac{\eta}{G\tau_d}
\end{equation}
is the ratio between the solvent viscosity and that of the quiescent
entangled solution.  Typical experimental values for the parameters
are $\epsilon\sim10^{-3}-10^{-5}$ and $Z\sim15-200$
\citep{adamsolmsted09a,tapadiawang06}.

As can be seen above, the choice $\delta=-1/2$ leads to a Newtonian
high shear rate branch, with slope dominated by the entanglement
contribution for typical values of $\epsilon$ and $Z$.  This is
inconsistent with the experiments of \citet*{tapadiawang03}, which
suggest a scaling $\sigma_{xy}\sim\dot{\gamma}^{1/2}$ at high shear
rate.  This scaling can be produced by choosing $\delta=1/2$
(Eq.~\ref{eqn:Tlimits}), which leads to
\begin{align}
\frac{\sigma_{xy}}{G} &= \left(\frac{\displaystyle3^{1/2}
    \dot{\gamma}\tau_d} {\displaystyle6
    Z\beta2^{1/2}}\right)^{\frac12} + \epsilon\dot{\gamma}\tau_d. 
\end{align}
This has an intermediate scaling $\sigma_{xy}\sim\dot{\gamma}^{1/2}$,
before giving way at the highest shear rates to Newtonian
behavior. However, this choice does a poorer job at the validation
tests carried out by \citet{likhtmangraham03} and has the incorrect
physical interpretation of the effect of stretch on CCR (as noted
above), and for these reasons we use $\delta=-1/2$ here. 

The Newtonian power law in the high shear rate branch, which applies
for our choice $\delta=-1/2$, produces a relatively narrow stress
plateau. This is one of the unsatisfactory approximations to the GLAMM
model that {may explain why such a large $Z$ value (compared to
  experiments) is required for the RP model} to exhibit velocity
recoil and transient banding.  This is an obvious and important
direction for future work.

\subsection{Constitutive Relations}
\begin{figure}[htb]
\includegraphics[width =0.48\textwidth]{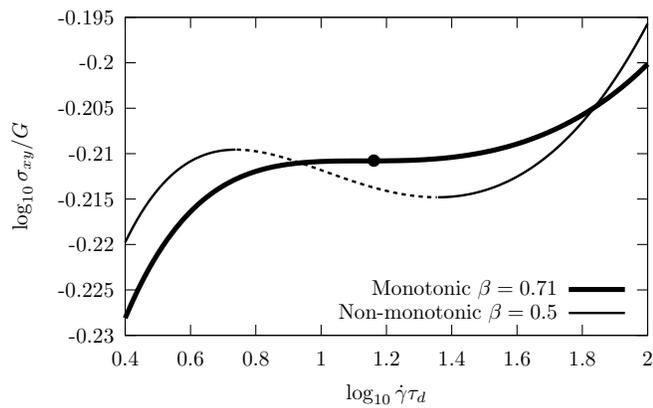}
\caption{Monotonic and non-monotonic constitutive curves arising from
  the Rolie-Poly model obtained by varying the CCR parameter
  $\beta$, for $\epsilon=10^{-5}$ and $Z=265$. The marked point at
  $\log_{10}\!\dot{\gamma}\tau_d=1.138\,(\dot{\gamma}\tau_d=13.74)$ on
  the monotonic 
  constitutive curve is studied in detail in
  Figs.~(\ref{fig:curvemeasures},\ref{fig:SPEV},\ref{fig:EVintegral}).}
\label{fig:CCs}
\end{figure}

The RP model readily admits non-monotonic constitutive curves that can
display shear banding, depending on the values of the CCR parameter
$\beta$, the entanglement number $Z$, and the solvent viscosity
$\epsilon$. For smaller $\epsilon$ and larger $Z$ there is a wider
separation between the high and low shear rate branches, while the CCR
parameter controls whether the crossover between the two regimes
contains a decreasing stress (smaller $\beta$ or less active CCR) or
is monotonic (larger $\beta$ or more active CCR). Fig.~\ref{fig:CCZs}
shows the effect of changing the number of entanglements for fixed
solvent viscosity and CCR parameter, while Fig.~\ref{fig:CCs}
illustrates the effect of varying the the CCR parameter.

\begin{figure}
\includegraphics[width =0.48\textwidth]{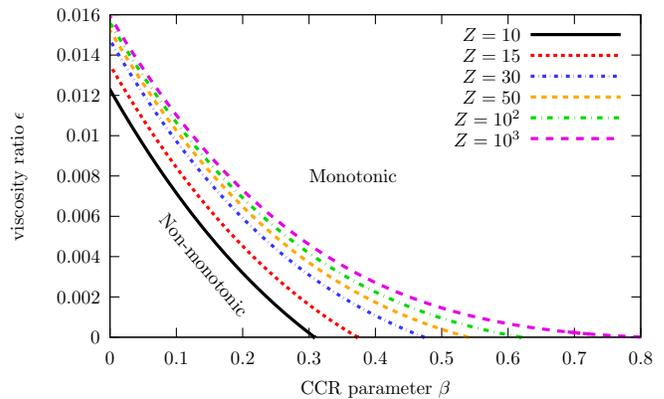}
\caption{Lines separating monotonic (upper right) from non-monotonic
  (lower left) constitutive behavior as a function of the viscosity
  ratio $\epsilon$ and CCR parameter $\beta$, for different
  entanglement numbers $Z$.}
\label{fig:paramsurface}
\end{figure}

The parameters space $(\beta, \epsilon, Z)$ for the Rolie-Poly model
can be divided into regions of monotonic and non-monotonic
constitutive curves, illustrated in
Fig.~\ref{fig:paramsurface}. Non-monotonic curves occur for small
values of $\epsilon$, small values of $\beta$, and large $Z$.

In what follows we will illustrate the transient banding-like behavior
associated with monotonic constitutive relations. We will study the
dynamics of startup to a shear rate that is on the plateau region, and
which eventually yields a non-banded homogeneous steady state.
Transient banding behavior develops under these conditions, and we
will explain this with the help of linear stability analysis.  We
primarily use the parameters $(\beta, \epsilon,Z) = (0.71,
10^{-5},265)$, corresponding to the thick line in Fig.~\ref{fig:CCs}.
Unfortunately, $Z=265$ is much higher than the values $Z\agt40-50$
needed to see shear banding experimentally. We believe that this is
because the Rolie-Poly model, as discussed above, only roughly
approximates the full microscopic theory of the GLAMM model, which is
itself an approximation; hence we caution the reader against
over-interpreting the values of these parameters.  We will return
later to briefly discuss this point.

\section{Dynamics and Boundary Conditions}
\subsection{Flow geometry}
We consider Couette flow between concentric cylinders of inner (outer)
radii $R_1$ ($R_2$), in cylindrical coordinates with flow varying only
in the radial direction, $\vec{v} = v(r) \hat{\thetabold}$. Using this
description the tangential and radial components of the force balance
conditions are respectively given by
\begin{eqnarray}
\label{eq:NStheta}
\rho \dot{v} &=& \frac{1}{r^2} \partial_r (r^2 \sigma_{r\theta})\\
- \rho \frac{v^2}{r} &=& \partial_r \sigma_{rr} + \frac{\sigma_{rr} -
  \sigma_{\theta \theta}}{r}, \label{eq:NS}
\end{eqnarray}
and the components of the polymer strain $\ten{W}$
obey
\begin{subequations}
  \begin{align}
    \!\!\mathcal{L} W_{\theta \theta}- 2 W_{r \theta}
    \dot{\gamma}  &= -\frac{2}{ \tau_R}(1-A)+\frac{2\mathcal{D}}{r^2}
    (W_{rr} - W_{\theta\theta})\\ 
    \!\!\mathcal{L}W_{rr} 
    &=-\frac{2}{ \tau_R}(1-A) + \frac{2\mathcal{D}}{r^2}
    (W_{\theta \theta} - W_{rr})\\ 
    \!\!\mathcal{L} W_{zz} &= -\frac{2}{ \tau_R}(1-A) \\
    \!\!\mathcal{L}W_{r\theta} -
    W_{rr}\dot{\gamma}&=\dot{\gamma} -
    \frac{4\mathcal{D}}{r^2}W_{r \theta} , 
  \end{align}
\end{subequations}
where $\dot{\gamma} = r\partial_r \left(\frac{v}{r}\right)$ is the
local shear rate and we have defined the nonlinear operator
\begin{equation}
\mathcal{L} = \partial_t - \frac{\mathcal{D}}{r} \partial_r
r \partial_r+ \frac{1}{\tau_d} + \frac{2}{\tau_R}
\left[1-A\left(\ten{W}\right)\right]\left[1+\beta
  A\left(\ten{W}\right)\right].
\end{equation}

The constitutive equations can be simplified by parametrizing
the spatial coordinate by 
\begin{align}
r &= R_1 e^{q y}, \label{eq:Rscale}\\
\noalign{\noindent where}
q &= \ln \frac{R_2}{R_1}\label{eq:q}
\end{align}
is a measure of the curvature of the measurement device, and the
dimensionless coordinate $y \in [0,1]$ spans the width of the gap $L=R_2-R_1$
\citep{greco97}.  

{The relative stress difference between the two cylinders is given by 
\begin{equation}
\frac{\sigma_{r\theta}(R_1)-\sigma_{r\theta}(R_2)}{\sigma_{r\theta}(R_2)}=1-e^{-2q}.
\end{equation}
A Couette cell with radii $R_1=2\,\textrm{cm},
R_2=2.1\,\textrm{cm}$ has $q=0.049$. In a 
cone and plate device the stress varies as \citep{LarsonComplex}
\begin{equation}
\sigma_{\theta\phi}=\frac{\sigma_{\theta\phi}(\theta=\pi/2)}{\sin^2\theta},
\end{equation}
where $\theta$ is measured with respect to the normal vector to the
plate. Hence the stress variation between the cone and plate is
\begin{equation}
\frac{\sigma_{\theta\phi}(\alpha)-\sigma_{\theta\phi}(0)}{\sigma_{\theta\phi}(0)}
= \tan^2\alpha, 
\end{equation}
where $\alpha$ is the cone angle. For cone angles
$\alpha=\pi/2-\theta\equiv4^{\circ}$ and $1^{\circ}$} the stress
difference is roughly equivalent to the values $q=2\times10^{-3}$ and
$q=2\times10^{-4}$, respectively \citep{BCAdams08}. {In what follows
  we will sometimes compare the stress variation of a cone and plate
  geometry to the equivalent stress variation of a Couette cell with a
  particular geometric parameter $q$: however, we do not perform
  calculations for the cone and plate geometry.}

We change to dimensionless quantities, labelled by $\tilde{}$, by
measuring time in units of the disengagement time $\tau_d$ and stress
in units of the plateau modulus $G$, so that the total
shear stress is expressed as
\begin{align}
 \tilde{\sigma}_{r \theta} &= W_{r\theta} +
 \epsilon\tilde{\dot{\gamma}}, \label{eq:shearrelation}
\end{align}
where $\tilde{\dot{\gamma}}=\dot{\gamma}\tau_d$ and
$\tilde{\sigma}_{r\theta} = \tilde{\sigma}_{r\theta}/G$. We define a
dimensionless diffusion constant
\begin{equation}
D \equiv \frac{\mathcal{D}\tau_d}{(q R_1)^2}.
\end{equation}
Note that the inner radius $R_1$ can be written in terms of the gap $L$ between cylinders as
\begin{equation}
R_1=\frac{L}{e^{q}-1},\label{eq:Rgap}
\end{equation}
so that $qR_1$ depends  only weakly on $q$ for small $q$.
\change{Below we will use $D$ and $q$ as independent parameters. Because we consider small $q\alt10^{-2}$, the dependence of $D$ on $q$ is negligible. Hence, by fixing $D$ and varying $q$ we can separate the effects of the total stress gradient, as parametrized by $q$, from the effects of the finite width $\sqrt{D}$ of the interface.}

In this representation the constitutive equation becomes 
\begin{subequations}
\label{eqn:allCC}
  \begin{align}
    \tilde{\mathcal{L}}  X &= - 6 (1-A)Z  +2 S \tilde{\dot{\gamma}} +
    2D e^{-2 q y} (Y-X)\label{eqn:XXCC}\\ 
    \tilde{\mathcal{L}} Y &= -6(1-A) Z
    - 2De^{-2 q y} (Y-X) \label{eqn:YYCC}\\
    \tilde{\mathcal{L}} W &=- 6 (1-A) Z  \label{eqn:ZZCC}\\
    \tilde{\mathcal{L}} S &= (1+Y) \tilde{\dot{\gamma}} - 4 D e^{-2 q y}
    S \label{eqn:SSCC},
  \end{align}
\end{subequations}
where $X=W_{\theta\theta}$, $Y=W_{rr}$, $W=W_{zz}$, and
$S=W_{r\theta}$, and $Z$ is the entanglement number given by
Eq.~(\ref{eq:entnumber}).  The nonlinear operator is thus
\begin{equation}
\tilde{\mathcal{L}} = \partial_{\tilde{t}} - De^{-2 q y}\partial_y^2+1 + 6Z(1-A)(1+\beta A),
\end{equation}
and the momentum balance equations become 
\begin{align}
\tilde{\rho} \dot{\tilde{v}} &= e^{-3 qy} \partial_y ( e^{2 q y}
\tilde{\sigma}_{r \theta})\label{eq:NSthetaB}\\
-\tilde{\rho} \tilde{v} &= \frac{1}{q} \partial_y \tilde{\sigma}_{rr} +
(\tilde{\sigma}_{rr} - \tilde{\sigma}_{\theta \theta}). 
\end{align}
where $\tilde{\rho} = \frac{\rho L^2}{G\tau_d^2}$. The dimensionless
velocity is scaled by $\tau_d$ and a length {$R_1(e^q-1)$ that becomes
  the plate separation $R_2-R_1=L$ in the $q\rightarrow0$ limit.

  Planar Couette flow obtains in the limit $q\rightarrow0$, with the
  correspondences} $\uvec{r}\rightarrow-\uvec{y}$,
$\widehat{\boldsymbol{\theta}}\rightarrow\uvec{x}$, $q\rightarrow 0$
and $qR_1 \rightarrow L$, and the force balance conditions become
$\tilde{\rho} \dot{\tilde{v}} = \partial_y ( \tilde{\sigma}_{y x})$
and $\partial_y \tilde{\sigma}_{yy}=0$. 
\subsection{Boundary Conditions}
\change{An imposed cylinder rotation rate $V/R_1$ can be expressed in
  terms of the integral of the shear rate across the gap by
\begin{equation}
\label{eqn:BC}
-\frac{V}{R_1} = \int_{R_1}^{R_2}  \dot{\gamma}\frac{dr}{r} =\int_0^1
\dot{\gamma} q\, dy
\end{equation}
The average applied shear rate is
  \begin{align}
    \label{eq:BCq}
    \langle{\dot{\gamma}}\rangle&\equiv \frac1q\int_{R_1}^{R_2}
    \dot{\gamma}\frac{dr}{r} = \int_0^1\dot{\gamma}\,dy.
  \end{align}
which reduces to $-V_1/L$ in the planar limit.}  We assume that there
is no slip at the walls, despite the fact that wall slip can be
important experimentally
\citep{Wan1999PCE,BouTap2006JR,boukany2009exploring}.

The presence of spatial gradients in the form of the `diffusion' term
necessitates a boundary condition on the viscoelastic strain
$\ten{W}$. Although it is possible to incorporate sophisticated wall
constitutive models \citep{BlaGra1996PRL} or complex boundary
conditions \citep{rossimckinley06,BCAdams08}, we will use the simplest
Neumann boundary conditions for the polymer strain,
\begin{equation}
(\boldsymbol{\hat{\textbf{n}}}\cdot\nabla) \ten{W} = 0\textrm{ for
} y = 0, 1;
\label{eqn:neumann}
\end{equation}
here $\boldsymbol{\hat{\textbf{n}}}$ is the normal to the boundary.

\section{Calculational Methods}
\subsection{Linear Stability Analysis}
\label{sec:LSA}
We first analyze the linear stability of Eqs.~(\ref{eqn:allCC}) above,
in the planar limit ($q\rightarrow 0$), \change{which should be a good
  approximation of the behavior of typical systems for which
  $q\simeq0.0001-0.05$.}  We follow the stability analysis performed
for the Johnson-Segalman model by \citet*{fielding03c}, who showed
that fluids with non-monotonic constitutive curves are unstable during
startup at fixed shear rate to spatial perturbations, which ultimately
develop into shear bands. We will apply this to monotonic constitutive
curves such as shown in Fig.~\ref{fig:CCs}.

\change{The equations of motion are first solved for a homogeneous
  time-dependent solution, which we refer to as the homogeneous base
  state. We  study the time dependence of inhomogeneous perturbations
  about this homogeneous base state. If these perturbations grow in
  time then the homogeneous time-dependent base state is unstable to
  inhomogeneous states, which we refer to as transient banding. Hence,
  we linearize the Navier-Stokes equation and the Rolie-Poly}
  constitutive equations around a uniform \change{time dependent}
  \change{base} state (denoted by subscripts 0), $X = X_0 + \delta X$,
  $Y = Y_0 + \delta Y$, $W = W_0 + \delta W$, $S = S_0 + \delta S$ and
  $\tilde{\dot{\gamma}} = \tilde{\dot{\gamma}}_0 + \delta
  \tilde{\dot{\gamma}}$. The linearized Navier-Stokes equation
  (Eq.~\ref{eq:NS}) is
\begin{eqnarray}
  \tilde{\rho} \partial_t \delta \tilde{\dot{\gamma}}
  &=&  \partial_y^2 \delta S +  
  \epsilon \partial_y^2 \delta\tilde{\dot{\gamma}},\label{eq:linearNS}
\end{eqnarray}
where $ \tilde{\dot{\gamma}}=\partial_y \tilde{v}$. Note that
$\tilde{\rho}$ is exceedingly small ($\tilde{\rho}\sim 10^{-9}$) for
the experiments of interest \citep{tapadiaravin06}, and justifies the
creeping flow limit in section \ref{sec:SR}. \change{However, we keep this
term in the linear stability analysis for completeness; leaving it out has no discernible effect for physically realistic values.}

We now consider spatial fluctuations in all quantities,
\begin{align}
\delta \vec{u}(y) &= \sum_{k} \delta \vec{u}_k e^{i k y},\\
\vec{u}&\equiv (X, Y, W, S, \tilde{\dot{\gamma}}),
\end{align}
where $\delta \vec{u}_k = \delta \vec{u}_{-k}^*$ because the function
$\delta \vec{u}(y)$ is real. The boundary conditions of
Eq.~(\ref{eqn:BC}) on the shear rate $\dot{\gamma}$ quantize the
values of the wavenumber, $k= n\pi\,(n=1,2,3,\ldots)$, to keep the
average shear rate across the gap fixed to the imposed
value. 

Linearizing Eqs.~(\ref{eqn:allCC},~\ref{eq:linearNS})  about the
homogeneous instantaneous state produces a matrix equation of the form
\begin{equation}
\partial_t \delta\vec{u}_k = \ten{M}(k,t)\cdot\delta\vec{u}_k, \label{eq:stab}
\end{equation} 
where the stability matrix is given by
 \\[5truept]
\begin{widetext}
{
\begin{align}
\ten{M} &=\left[\begin{matrix}
\Pi - ZA^3(1+ \Gamma  X)  & - A^3 Z (1+ \Gamma  X)
& - A^3 Z (1+ \Gamma  X)&2\tilde{\dot{\gamma}}&2S\\[7truept]
- A^3 Z (1+ \Gamma  Y) & \Pi - ZA^3(1+ \Gamma  Y)  
& - A^3 Z (1+ \Gamma  Y)&0 & 0 \\[7truept]
- A^3 \Gamma Z W & - A^3 Z (1+ \Gamma  W) & 
\Pi - ZA^3(1+ \Gamma  W)  & 0 & 0 \\[7truept]
 -A^3 \Gamma Z S&  -A^3\Gamma W S+ \tilde{\dot{\gamma}}
 &-A^3 \Gamma Z S&\Pi &1+Y \\[7truept]
0&0&0&-\frac{k^2}{\tilde{\rho}} & -\frac{k^2\epsilon}{\tilde{\rho}}
\end{matrix}
\right],
\end{align}
}
\end{widetext}
{with
\begin{align}
\Pi&=-1 -D k^2 - 6 Z (1-A) (1+A \beta)\\
\Gamma &= 1 - \beta + 2 A \beta.
\end{align}
}

The matrix depends on the values of the shear rate and
polymer strain at a time $t$, determined according to  homogeneous
evolution of the dynamics.  A
linear disturbance $\delta\vec{u}_k$ will grow exponentially in
time if an eigenvalue $\omega_{\alpha}$ of $\ten{M}$ has
positive real part \citep{fielding03c}.
We define $\omega_{\textrm{max}}$ as the
eigenvalue with the largest real part $\Re(\omega_{\alpha})$.  To
analyze the stability of the transients with respect to spatial
fluctuations we calculate the evolution of the eigenvalues of
$\ten{M}$ as a function of time $t$, using the homogeneous solution
$\ten{W}(t)$ for the unperturbed state at a given time.

A useful measure of the duration and severity of instability is the
total `weight' $\Omega_{\textrm{max}}$ of the instability, obtained by
integrating the real part of $\omega_{\textrm{max}}$ over all time
during which it is positive:
\begin{equation}
\Omega_{\textrm{max}} = \int_0^\infty \textrm{Max}(0, \Re(\omega_{\textrm{max}}) ) dt.
\end{equation}
For larger weights $\Omega_{\textrm{max}}$ we expect the fluid to
enter the non-linear regime and develop a transient shear band, after
which stable homogeneous flow is restored in steady state (for
monotonic constitutive curves). This is verified by the full
non-linear calculation, to which we now turn.

\subsection{Spatially resolved Rolie-Poly model}
\label{sec:SR}
We have thus far addressed the evolution of the homogeneous model, and
the linear (in)stability of this evolution to inhomogeneous states.
To compute the full inhomogeneous non-linear dynamics we solve the
creeping flow equation, Eq.~(\ref{eqn:creep}), and the DRP
constitutive equations, Eqs.~(\ref{eqn:allCC}), with a constrained
shear rate, Eq.~(\ref{eqn:BC}) and the boundary conditions,
Eq.~(\ref{eqn:neumann}). This leads to a set of coupled second order
partial differential equations. A uniform spatial grid of $500$ points
was used to discretize the equations, and a semi-implicit
Crank-Nicolson scheme was used to effect the time evolution. We have
checked convergence with respect to timestep and spatial mesh. Typical
dimensionless time steps were $10^{-5}$, with dimensionless spatial
meshes of $2\times10^{-3}$; and a dimensionless diffusion constant
$D=4\times10^{-4}$ was used, corresponding to an effective
dimensionless diffusion length $2\times10^{-2}$. 

The shear rate constraint of Eq.~(\ref{eqn:BC}) was implemented by
integrating Eq.~(\ref{eq:NSthetaB}) in the zero Reynolds number limit
\change{to obtain}
\begin{subequations}
\label{eq:torque}
  \begin{align}
    T e^{-2qy} &= \tilde{\sigma}_{r \theta} = W_{r\theta} +
    \epsilon\tilde{\dot{\gamma}},\label{eq:gamelim1}\\
    T &= \frac{2q}{1-e^{-2q}}\left[\left\langle{W}_{r\theta}\right\rangle +
        \epsilon\left\langle\dot{\gamma}\right\rangle\right], 
\label{eq:gamelim2}  
  \end{align}
\end{subequations}
\change{where $T$ is the torque per unit length
  applied to the inner cylinder.}  \change{Eqs.~(\ref{eq:torque}) were
  then used to eliminate the local shear rate
  $\tilde{\dot{\gamma}}(y)$ in Eqs.~(\ref{eqn:allCC}) in terms of the
  torque and the  local value of the polymer shear strain
  $W_{r\theta}(y)$:
\begin{align}
  \tilde{\dot{\gamma}}(y) &= \frac{1}{\epsilon}
  \left[T e^{-2qy} - W_{r\theta}(y)\right]. 
\end{align}
The torque per unit length $T$ can be computed at any instant in time
from the full inhomogeneous profile in terms of the spatial averages
of $W_{r\theta}$ and $\tilde{\dot{\gamma}}$, as defined in
Eq.~(\ref{eq:gamelim2}).}

There are several possible measures of the degree of inhomogeneity in
the spatially resolved flow profile. At any instant we define the shear rate drop
\begin{align}
\Delta &=
\frac{|\dot{\gamma}_\textrm{max} - \dot{\gamma}_\textrm{min}|}
{\langle{\dot{\gamma}}\rangle}, \label{eq:Delta}   
\end{align}
where $\dot{\gamma}_{\textrm{max}}$
(resp. $\dot{\gamma}_{\textrm{min}})$ is the maximum (resp. minimum)
of the measured shear rate profile, which should be easily accessible
from experimental data. Other measures could include the variance of
the shear rate values across the sample cell, or the maximum local
gradient of the shear rate. Since all of these give an equivalent
qualitative indication of the development of inhomogeneities during
flow, we choose this simple measure.
 
\begin{figure}[htb]
\includegraphics[width = 0.48\textwidth]{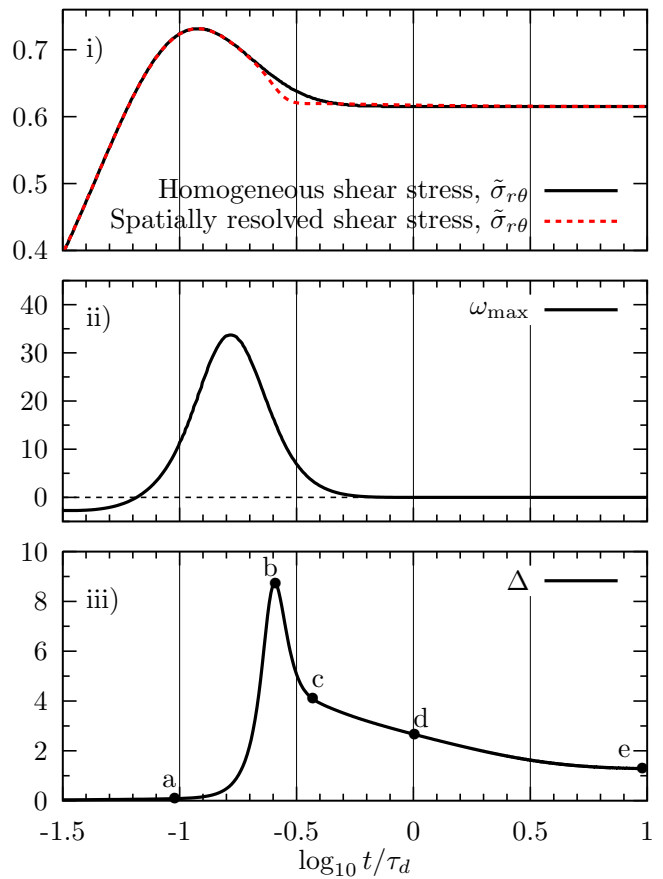}
\caption{Evolution of various quantities as a function of time during
  shear rate startup for parameters
  $(\beta,\epsilon,Z)=(0.71,10^{-5},265)$, for $q=10^{-3}$ and applied
  average shear rate $\dot{\gamma}\tau_d=13.74\,
  (\log_{10}\dot{\gamma}\tau_d=1.138)$ (the marked point on the
  monotonic constitutive curve in Fig.~\ref{fig:CCs}). (i) Dimensionless total shear 
  stress calculated based on homogeneous flow
  (solid) or allowing inhomogeneous
  transients (dashed); (ii) real part of the largest unstable
  eigenvalue $\omega_{\textrm{max}}$; (iii) shear rate drop $\Delta$
  (as defined in Eq.~\ref{eq:Delta}), as a measure of spatial
  inhomogeneity showing significant transients during startup. The
  times \texttt{a}, \texttt{b}, \texttt{c}, \texttt{d}, \texttt{e} are
  shown as velocity profiles in Fig.~\ref{fig:SPEV}.}
\label{fig:curvemeasures}
\end{figure}
\section{Results: instability and transients}
In this section we study the linear (in)stability of homogeneous
startup, and compare this with the non-linear spatially-resolved
calculations in which the transient shear bands are allowed to form
naturally. Here we consider an initial state at rest with no `noise',
in a weakly curved geometry $q=10^{-3}$. In
Sec.~\ref{sec:role-pert-trans} we will study the effects of different
magnitudes of spatial noise and curvature.

We study model parameter values
$(\beta,\epsilon,Z)=(0.71,10^{-5},265)$, corresponding to the solid
constitutive curve in Fig.~\ref{fig:CCs}, for an imposed shear rate
$\dot{\gamma}\tau_d=13.74$ (the marked point at
$\log_{10}\dot{\gamma}\tau_d=1.138$) and with shear stress gradient
parametrized by $q=10^{-3}$.  
  During startup the shear stress overshoots, before
settling down to steady state on a time of order
$\tau_d$. {Fig.~\ref{fig:curvemeasures} compares a calculation in
  which the shear rate remains homogeneous (solid line) with a
  spatially-resolved calculation that allows for inhomogeneities
  (dashed line), showing that the stress decays more rapidly in the
  spatially-resolved model.} The largest unstable eigenvalue, from the
linear stability analysis, becomes positive for
$\log_{10}(t/\tau_d)\agt-1.2$, and reaches a maximum shortly after the
stress overshoot. At late times it decays to a slightly negative
value, controlled by the very shallow slope of the steady state
constitutive curve.
\begin{figure}[!htb]
\includegraphics[width = 0.48\textwidth]{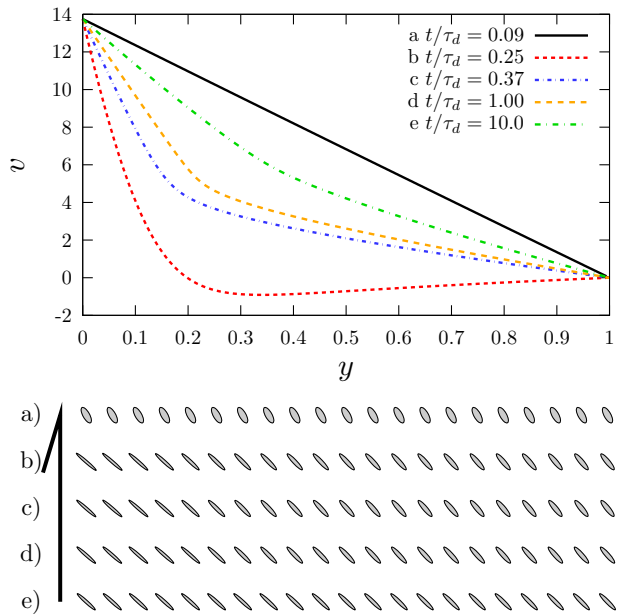}
\caption{Velocity profiles during startup flow  for the
  parameters of the marked point
  $\log_{10}\dot{\gamma}\tau_d=1.138\,\,(\dot{\gamma}\tau_d=13.74)$ on
  the monotonic constitutive curve in Fig.~\ref{fig:CCs}:
  $(\beta,\epsilon,Z)=(0.71,10^{-5},265)$, for $q=10^{-3}$. The
  corresponding ellipsoids indicating the components of the
  $W_{rr}$, $W_{\theta\theta}$ and $W_{r \theta}$
  components of the polymer strain are shown below. At time
  $t=0.25\tau_d$, which is shortly after the stress overshoot, both
  negative recoil velocities and transient shear banding occur. The
  times labeled \texttt{a}, \texttt{b}, \texttt{c}, \texttt{d},
  \texttt{e} are the points labeled on $\Delta(t)$ in
  Fig.~\ref{fig:curvemeasures}.}
\label{fig:SPEV}
\end{figure}

Associated with this positive eigenvalue  is a pronounced inhomogeneous
transient during startup, as evidenced by the shear rate drop
$\Delta$ in Fig.~\ref{fig:curvemeasures} and the velocity profiles in
Fig.~\ref{fig:SPEV}. The most significant heterogeneity comes shortly
after the eigenvalue has reached its maximum, after the
instability has had sufficient time to develop non-linear
consequences.  Fig.~\ref{fig:SPEV} shows that this maximum (at time
$t=0.25\tau_d$, labeled \texttt{b}) corresponds to a dramatic
transient banding profile, in which the velocity field has
\text{recoiled} and become negative at the outer part of the cell
(closer to $y=1$), where the total stress is slightly lower. The
strong transient bands decay after a few $\tau_d$ until the steady
state profile is reached. In this case, the stress gradient
corresponding to $q=10^{-3}$ is enough to induce significant
heterogeneity even in steady state \citep{adamsolmsted09a} (we will
return to the role of the stress gradient in
Sec.~\ref{sec:role-pert-trans}). Fig.~\ref{fig:SPEV} also shows the degree
of polymer deformation (strain) during the recoil. The polymer strain
tensor becomes well-ordered, at an angle of approximately $45^\circ$
with respect to the flow direction, when the transient develops
(\texttt{b}). Interestingly, it remains ordered and changes only
weakly until steady state is reached; moreover, the spatial gradient
only introduces a relatively small gradient in the molecular order and
orientation. Hence, the transient inhomogeneity would be difficult to
observe using a molecular probe such as birefringence or neutron
scattering, as compared to explicit measurements of velocity profiles.

\begin{figure}
\includegraphics[width = 0.48\textwidth]{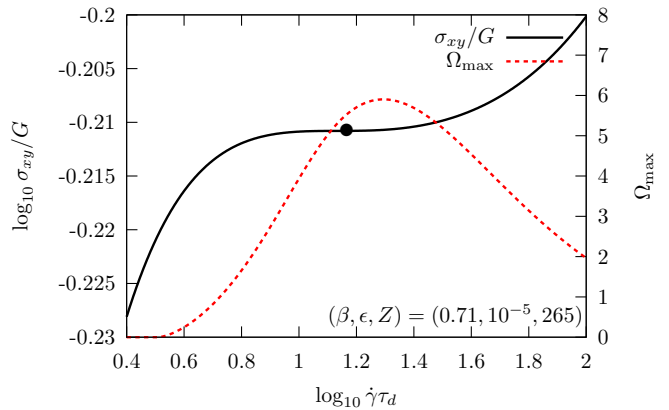}
\caption{The weight $\Omega_{\textrm{max}}$ of the instability
  (magnitude of the integral of the most unstable eigenvalue) and the
  monotonic constitutive curve $\sigma_{xy}(\dot{\gamma})$ (solid line)
  from Fig.~\ref{fig:CCs}. For $(\beta,\epsilon,Z)=(0.71,10^{-5},265)$,
  for $q=10^{-3}$. }
\label{fig:EVintegral}
\end{figure}
Having explored the response to a single shear rate startup protocol,
we now study the strength of the instability for different shear
rates.  Fig.~\ref{fig:EVintegral} shows the weight
$\Omega_{\textrm{max}}$ of the instability (integrated unstable
eigenvalue) for all applied shear rates for this parameter set. The
largest weight occurs near the flattest part of the stress plateau,
and indeed we find the clearest transient banding in this region. A
detailed study (not shown here) reveals that the weight
$\Omega_{\textrm{max}}$ is larger for large $Z$, small $\epsilon$ and
larger $\beta$; \textit{i.e.} parameters for which the two flow
branches are more widely separated and the constitutive curve has a
shallower slope. The larger value of $\beta$ produces a flatter,
longer stress plateau.

In summary, transient banding is initiated by a linear instability
that begins during startup, and becomes largest shortly after the
stress overshoot. If this instability has enough time and strength to
grow, as parametrized by the weight $\Omega_{\textrm{max}}$, then
transient banding and even recoil can appear before the eventual
steady state is reached. The steady state is not banded, but weakly
inhomogeneous as specified by the stress gradient across the
rheometer.
\section{The role of perturbations in 
transient banding}\label{sec:role-pert-trans}
We have seen thus far that inhomogeneous flow, or equivalently
transient shear banding, is correlated with the most unstable
eigenvalue governing fluctuations away from a homogeneous state at a
given instant in time. We now study the spatial perturbations needed
to trigger the instability, by evolving the non-linear inhomogenous
equations of motion.  There are several sources of spatial
inhomogeneity that could serve this purpose: curved rheometer devices
(cone and plate or cylindrical Couette geometries) possess a spatial
inhomogeneity in the shear stress; thermal or instrument noise is
presumably always present; thermal gradients across the cell may
persist; and there may be inhomogeneous initial conditions due to very
slow relaxation after sample loading. Here we will study (i) the
consequences of a stress gradient due to curved streamlines and (ii)
the role of inhomogeneous initial conditions that may arise due to the
other effects just mentioned.

\begin{figure*}[tbh]
\includegraphics[width = 0.95\textwidth]{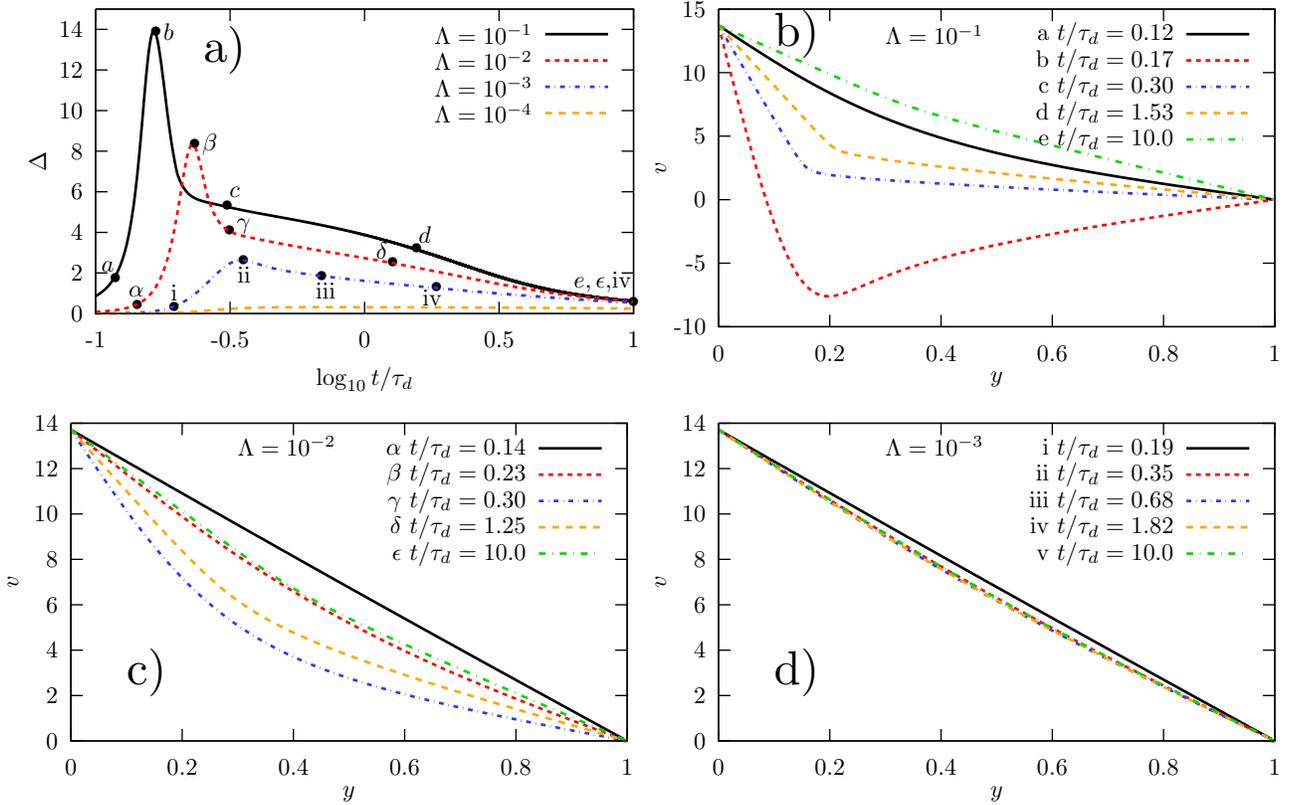}
\caption{Effect of noisy initial conditions for $W_{yy}$ on (a)
  the shear rate drop $\Delta$ as a measure of inhomogeneity, and
  (b-d) transient velocity profiles $v(y)$, for different initial
  conditions $W_{yy}=\Lambda\cos \pi y$. In steady state
  $W_{yy}=-0.52$. For parameters
  $(\beta,\epsilon,Z)=(0.71,10^{-5},265)$ and $q=0$.}
\label{fig:noisyIC}
\end{figure*}
\begin{figure}[tbh]
 \includegraphics[width = 0.48\textwidth]{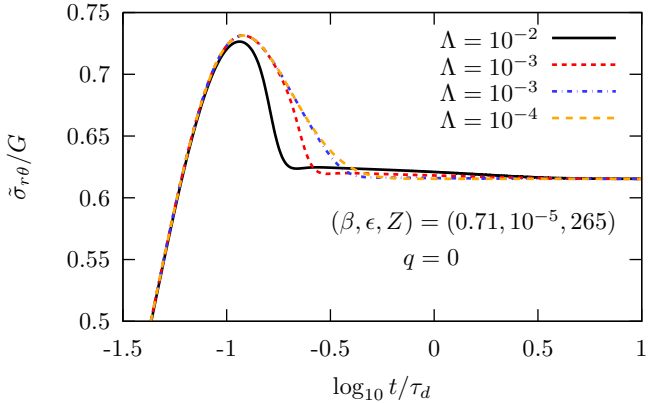}
 \caption{Stress overshoots for initial conditions conditions
   $W_{yy}=\Lambda\cos \pi y$, parameterized by $\Lambda$. For
   parameters $(\beta,\epsilon,Z)=(0.71,10^{-5},265)$.}
\label{fig:ICperturbovershoot}
\end{figure}

We first consider the effect of random initial conditions in a
geometry with no intrinsic stress gradient (\text{i.~e.} planar Couette
flow, $q=0$). A cosine wave with wavelength $2 L$, in keeping with the
boundary conditions, was set as the initial condition for a given
polymer strain component, with the others set to zero. The
constitutive equations were then evolved and the velocity profile
calculated. This was carried out for a range of different amplitudes
$\Lambda$ of the cosine wave. The different components of the polymer
strain have markedly different effects: {perturbations in $W_{yy}$
  have the most dramatic effect, because this strain component (along
  the flow gradient) is advected into the polymer shear strain
  $W_{xy}$ (Eq.~\ref{eqn:SSCC}), which in turn directly contributes to
  the measured total shear stress. Conversely, perturbations in
  $W_{xy}$ are much less important because they are rotated into the
  velocity direction ($W_{xx}$), which does not contribute to the
  total shear stress.} Fig.~\ref{fig:noisyIC} shows the effect of an
initial condition in $W_{yy}$. For small enough initial noise there is
virtually no transient banding, while substantial initial noise (a few
percent of the steady state polymer strain in flow) can lead to strong
transient banding and negative velocity recoil. {The modulus of
  entangled polymers is $G\simeq ck_BT$, where $c$ is the
  concentration of entanglement strands. Reasoning that coherent
  fluctuations can obtain within a volume of order $a^3$, where $a$ is
  the tube diameter, \citet{marrucci1983free} estimated the typical
  average strain $\langle \delta W_{\alpha\beta}\rangle$ due to
  thermal fluctuations in entangled polybutadiene melts to be of order
  $\sqrt{1/6.4}\simeq0.39$, assuming $a\simeq5.8\,\textrm{nm}$}; this
is comparable to the initial condition necessary to induce a transient
banding response in our numerical calculation. 

\change{The stress overshoot also changes character
  (Fig.~\ref{fig:ICperturbovershoot}): for a larger value of $\Lambda$
  (a stronger perturbation) the stress relaxes more quickly after the
  overshoot, due to the more rapid development of transient
  banding. For $\Lambda=10^{-2}$ the stress actually
  \textit{undershoots} the steady state value before increasing and
  then finally decreasing slowly to the steady state. Similar stress
  undershoots were reported by \citet{Tapadia2004Nonlinear-flow-} and
  by \citet*{sui2007iep}, for entangled polymer solutions that
  underwent flow instability; and by \citet{crawley1977geometry} in
  much earlier work.}

\begin{figure*}[tbh]
 \includegraphics[width = 0.95\textwidth]{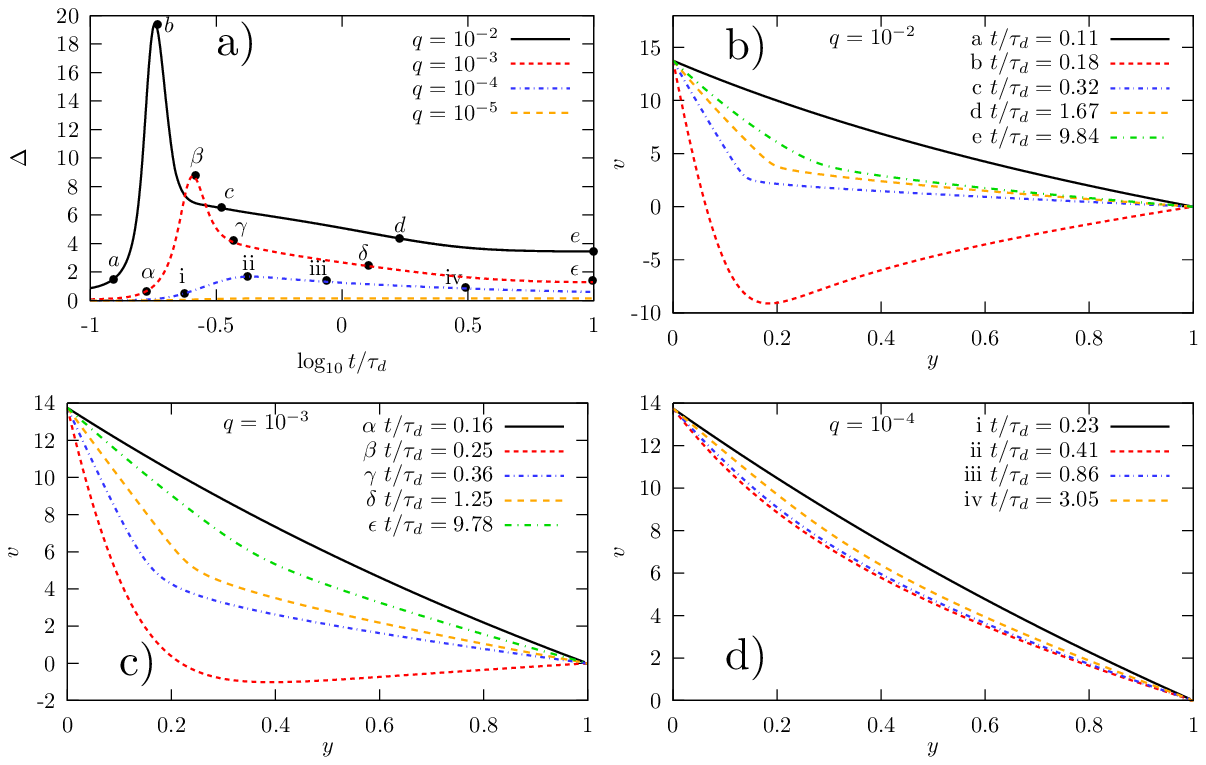}
 \caption{Effect of different stress gradients
   $q\equiv\ln\frac{R_2}{R_1}$ on (a) the shear rate drop $\Delta$ as
   a measure of inhomogeneity, and (b-d) transient velocity profiles
   $v(y)$. For parameters
   $(\beta,\epsilon,Z)=(0.71,10^{-5},265)$. The value $q=10^{-3}$
   corresponds roughly to a cone angle of $2.5^{\circ}$.}
\label{fig:qperturb}
\end{figure*}

\begin{figure}[tbh]
 \includegraphics[width = 0.48\textwidth]{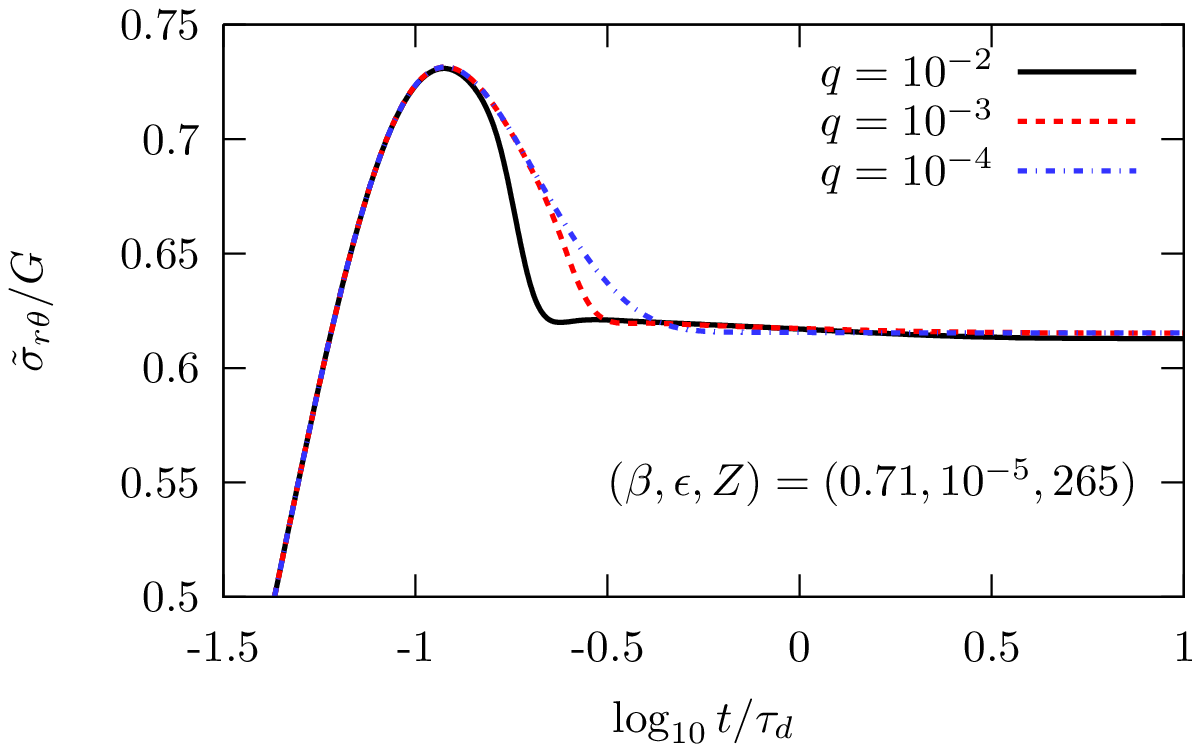}
 \caption{Stress overshoots for different stress gradients
   $q=\ln\frac{R_2}{R_1}$. For parameters
   $(\beta,\epsilon,Z)=(0.71,10^{-5},265)$.}
\label{fig:qperturbovershoot}
\end{figure}

An alternative perturbation is a stress gradient due to,
\textit{e.g.}, the curvature of a Couette cell or a non-zero angle in
a cone and plate rheometer. This is studied in
Fig.~\ref{fig:qperturb}. As explained earlier we parametrize this by
the curvature parameter $q=\ln R_2/R_1$ (Eq.~\ref{eq:q}). For
$q>10^{-3}$, whose stress gradient is similar to that of very thin gap
Couette rheometers or typical cone and plate rheometers, there is
substantial transient banding for these parameters; while for
$q=10^{-5}$ there is essentially no transient banding. \change{[We
  remind the reader that this calculation was performed for
  cylindrical Couette flow, and that the comparison with cone and
  plate flow is based solely on the shear stress gradients in the two
  geometries.] The stress overshoot also relaxes quicker after the
  overshoot for a larger stress gradient (larger $q$)
  (Fig.~\ref{fig:qperturbovershoot}).}

An experimental test is thus to observe the change in transient
banding as a function of flow geometry and hence $q$; or to
systematically pre-shear the material to induce molecular deformation;
or otherwise induce or spatially inhomogeneous initial conditions into
a sliding plate geometry. Existing experiments show transient banding
in both curved and flat geometries; our calculations would be
consistent with reproducible transients in a curved geometry, but less
reproducible transients in a flat geometry, reflecting the variations
of initial and loading conditions.

\section{Discussion}
\subsection{Parameter Space and Critique of the Rolie-Poly Model}
\begin{figure*}[tbh]
\includegraphics[width = 0.98\textwidth]{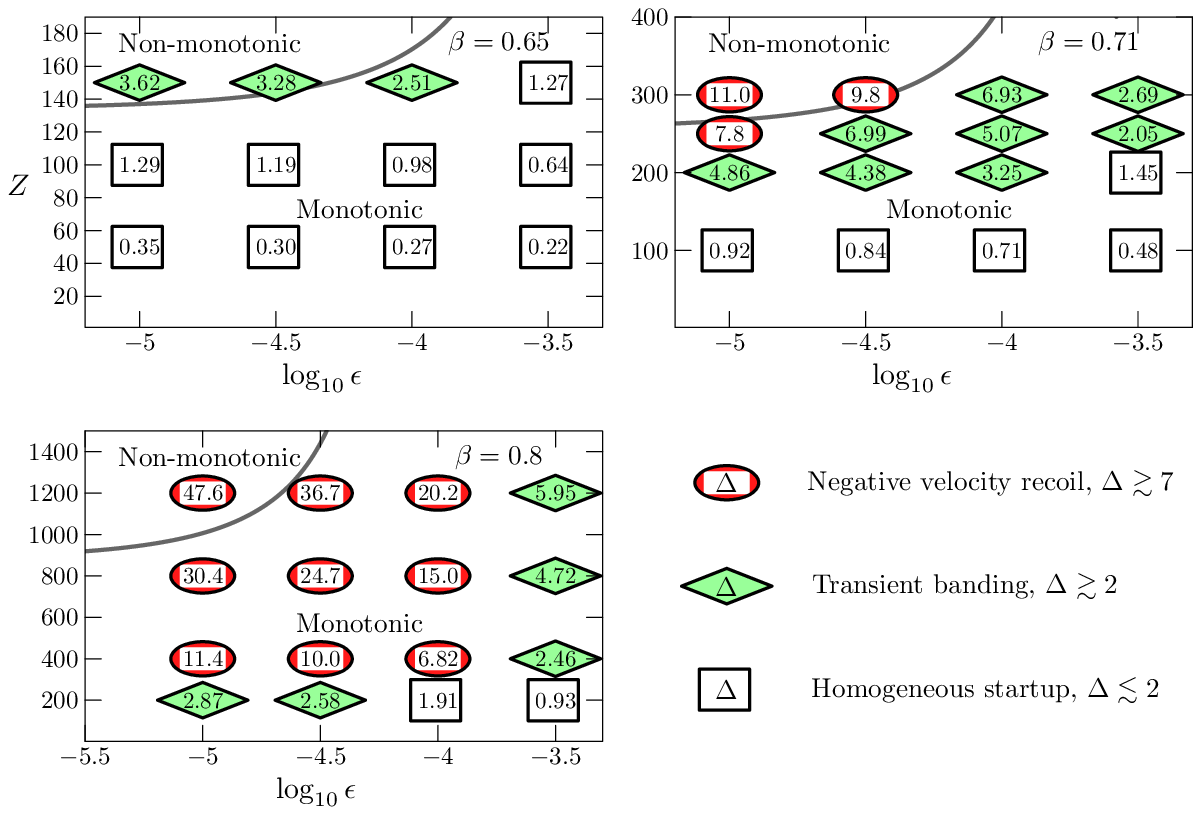}
\caption{Regions of parameter space in which one expects to find
  different phenomena, for $\beta=0.65, 0.71, 0.8$ and
  $q=10^{-3}$. The numbers in boxes denote the largest value of the shear rate drop
  $\Delta$ found during startup for an applied average shear rate on the
  flattest part of the steady state constitutive curve, analogous to
  the solid circle in Fig.~\ref{fig:EVintegral}.  Shown are: steady
  state banding in the non-monotonic region; transient banding
  ($\Delta\agt2$ and diamonds or ellipses); and negative velocity
  recoil upon startup ($\Delta\agt7$ and ellipses). }
\label{fig:parameterspace}
\end{figure*}

We have explored the behavior of transient banding as a function of
the parameters of the DRP model (Fig.~\ref{fig:parameterspace}). A
monotonic constitutive curve is more (dynamically) unstable and
susceptible to transient banding if it has well separated flow
branches, which occurs for greater numbers of entanglements $Z$,
larger convective constraint release (CCR) efficiency $\beta$, and
lower solvent viscosity $\epsilon$. The larger CCR parameter $\beta$
may appear paradoxical, since increasing it renders a banding fluid
stable. However, one also needs, simultaneously, a smaller viscosity
ratio and larger $Z$, to render the plateau shallower and wider in the
monotonic regime. Although the value of the CCR parameter $\beta$ is
evidently important, we have as yet no way of independently
determining it.  \citet{likhtmangraham03} used the values of $\beta =
0, 0.5$ and $1$ to fit multimode, steady state and transient data
respectively. It would be useful to compare a quantitative model such
as the DRP model with experimental results on shear banding (transient
or steady state) to try and infer the magnitude of the CCR parameter,
given knowledge of the other parameters.

Unfortunately the DRP model is still too crude to make this a
justifiable exercise.  The high shear rate branch is not handled
correctly, since Newtonian behavior is not observed experimentally at
the highest shear rates \citep{tapadiawang03}. Moreover, the DRP model
is only a simple approximation to the more complete GLAMM theory
\citep{likhtmangraham03,graham2003microscopic}, and there is thus not
a precise correspondence between the parameter values used to fit the
DRP model to data, and their microscopic physical meaning in the
original GLAMM model.  For example, we use rather larger values of
entanglement number $Z$ than are realized in most
experiments. However, the DRP model reproduces the qualitative
behavior, albeit without the quantitative material parameters. These
qualitative conclusions seem to be robust, as far as we can discern:
the much broader and flatter steady state constitutive curves are
accompanied by dramatic dynamics that can result in transient
inhomogeneities. There is an urgent need for performing inhomogeneous
calculations with more detailed theories such as the GLAMM
model. \change{The GLAMM model itself is also only an approximation,
  and it  certainly  misses important physics about the
  behaviour of the tube at very high shear rates
  \citep{grahammcleish2007}.}

Fig.~\ref{fig:parameterspace} shows the behavior expected for three
values of $\beta$, as a function of entanglement number $Z$ and
viscosity ratio $\epsilon$, for $q=10^{-3}$. The figure shows regions
of (1) steady state banding signified by a non-monotonic constitutive
curve; (2) transient banding during startup; and (3) dramatic recoil
with negative velocities during startup. From our startup calculations
we found these phenomena to occur for shear rate drops (see Eq.~\ref{eq:Delta}) 
 $\Delta\agt2$ (transient
banding) and $\Delta\agt7$ (transient banding featuring srecoil).
Experiments show transient banding for $Z\sim 15-50$, steady state
banding for higher $Z$, and recoil for some fluids that do not shear
band in steady state. Based on this, we suggest that the experimental
phenomenology of entangled polymer solutions most resembles that for
values of the CCR parameter $\beta$ in the range between $\beta=0.71$
and $\beta=0.8$, \textit{as parametrized by the Rolie-Poly model}.  We
emphasize that the parameter values for this model do not necessarily
correspond to those of experiments. However, polymer solutions have a
reasonably wide range of $Z$ for which transient, and not steady
state, banding is found as exhibited in Fig.~\ref{fig:parameterspace}
for $\beta=0.71$.

\subsection{The stress overshoot and the non-monotonicity of the
  instantaneous constitutive curve}
Fig.~\ref{fig:curvemeasures} shows that transient banding occurs
during the decreasing stress after the stress overshoot, and occurs a
short time after the unstable eigenvalue is most unstable. At these
shear rates the fluid behaves elastically at early times, and the
stress overshoot as a function of time could thus be envisioned as the
stress overshoot of a \textit{solid} as a function of strain
$\gamma=\dot{\gamma}t$; such an elastic material with $\partial
\sigma_{xy}/\partial\gamma<0$ could be expected to be unstable
\citep{marrucci1983free}. This was also noted by \citet*{sui2007iep}
in their study of entangled polymer solutions

\begin{figure}
\includegraphics[width=0.48\textwidth]{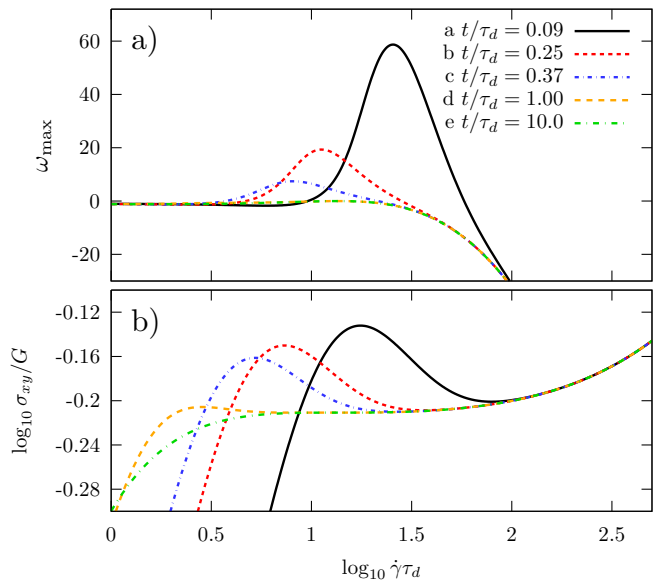}
\caption{(a) Maximum eigenvalue as a function of shear rate
  $\dot{\gamma}$ for different times $t$ after startup. (b)
  instantaneous constitutive curves, calculated from starting up at
  different shear rates $\dot{\gamma}$. Each curve represents the
  stress obtained a time $t$ after starting up at the given shear rate
  $\dot{\gamma}$. Note the strong correlation between the region of
  positive maximum eigenvalue, and the region of negative slope of the
  time dependent constitutive curve.}\label{fig:instantaneous}
\end{figure}

Fig.~\ref{fig:EVintegral} shows that the integrated instability, as
determined by the linear stability analysis, is most pronounced near
the flattest part of the constitutive curve. Since the steady state is
nearly unstable at this point, it is tempting to declare the nearly
flat slope as the cause of the instability. However, the steady state
behavior may have little influence on the instantaneous dynamics. To
explore this, we have calculated the instantaneous constitutive curve,
as described in the introduction and constructed by
\citet{hayes2010constitutive} from their experimental data. We
calculate the startup flow for different \textit{homogeneous} shear
rates to obtain a surface $\sigma_{xy}(\dot{\gamma},t)$. This surface
then defines a curve relating shear stress to shear rate at any
instant in time $t$ (the \textit{instantaneous constitutive curve}).
If this curve has a negative slope, then one expects an instability in
the dynamics at finite Reynolds number, or at least in a subspace of
the full dynamics specified by the matrix $\ten{M}$ that appears in
the linear stability calculation of
Section~\ref{sec:LSA}. Fig.\ref{fig:instantaneous} shows that the
instantaneous constitutive curve becomes non-monotonic, with the
strongest instability (largest unstable eigenvalue
$\omega_{\textrm{max}}$) indeed occurring very close the point of
steepest negative slope $\partial \sigma_{xy}/\partial\dot{\gamma}$.

Hence, transient banding, while evidently linked to the steady state
constitutive curve, may be expected when the instantaneous
constitutive curve is unstable (has negative slope), based on linear
stability analysis. It may also be expected during a stress overshoot,
for applied shear rates that approximate a step strain and are thus
strong enough to bring the fluid into its elastic regime
\citep{marrucci1983free}. We emphasize that we have shown a very
strong correlation, but in the context of the Rolie-Poly model. The
precise correspondences remain to be explored in other models, and we
await more work on this subject.

\subsection{Other transient phenomena}
Two other transient banding phenomena have been reported. Well defined
shear bands have been observed during portions of the cycle during
Large Amplitude Oscillatory Shear (LAOS) in polymer solutions
\citep{tapadiaravin06,ravindranath2008large}. This has also been
calculated numerically by \citet{adamsolmsted09a} using the DRP model
for a monotonic constitutive curve, and by \citet{zhouLAOS2010} for a
non-monotonic constitutive model devised for wormlike micelles. The
calculations revealed banding-like transients for shear rates sweeping
across the pseudo-plateau of a barely monotonic constitutive model,
and for frequencies slower than the inverse reptation time. This is
slow enough for a transient band to form during the upward sweep of
strain rate (according to Fig.~\ref{fig:curvemeasures}), but not so
slow that it decays away completely. It can then be reinforced in
subsequent sweeps, given the strong spatial perturbation it
represents, until it develops a sharp profile after a few cycles.
 
In other experiments, step strain at shear rates
$\dot{\gamma}\tau_R>1$ leads to an inhomogeneous response (including
negative velocity recoil) after cessation of step strain, in both
solutions \citep{wang2006nonquies,WangMM2007} and melts
\citep{boukany2009step}. In one example, a solution of $Z=64$ was
strained to just the stress maximum in the overshoot. During
relaxation, recoil and an inhomogeneous response developed, similar to
our previous calculations \citep{adamsolmsted09b}.  One explanation
for this result is that the fluid became unstable during strain, and
the instability was able to amplify a nascent fluctuation large enough
to induce a non-linear and heterogeneous response during the
subsequent relaxation. Indeed, the calculation in
Fig.~\ref{fig:curvemeasures} shows that the largest eigenvalue can
become positive, signifying instability, before the overshoot is
reached. However, this cannot be the entire story. In melts the
inhomogeneous response can also be a dramatic fracture
\citep{boukany2009step} that, as yet, we cannot calculate from the DRP
model \citep{wangcomment2009,adamsolmsted09b}.  This could be because
  the DRP model inadequately captures the physics of the GLAMM model;
  or because important physics governing the heterogeneity associated
  with retraction has yet to be correctly incorporated in any model
  \citep{wang07}.

\subsection{Implications for validating constitutive models}
A significant amount of literature, both experimental and theoretical,
has been devoted to testing constitutive models against the initial
stress transients, including the overshoot and subsequent relaxation
\citep{likhtmangraham03,wang2006nonquies}. This is typically done by
comparing the stress transient for shear rate startup to a calculation
that assumes homogeneous shear flow in a planar geometry. As is
evident in Fig.~\ref{fig:curvemeasures}, the relaxation after the
overshoot is faster when the fluid is allowed to become inhomogeneous,
as physically occurs during transient banding. Cone and plate
geometries are often used because unlimited strains can be applied and
the weak stress gradient is usually hoped to be negligible.
Unfortunately, the transient banding and hence its associated stress
response can be pronounced even for very small stress varations, as
parametrized by $q$. A planar geometry should exhibit less transient
banding if there are no significant perturbations
(Fig.~\ref{fig:qperturb}). We are not yet able to quantify this level
of perturbation.
 
Another common benchmark is the Doi-Edwards damping function
$h(\gamma,t)\equiv \sigma_{xy}(\gamma,t)/\sigma_{xy}(0,t)$, defined as
the stress relaxation $\sigma_{xy}(\gamma,t)$ after a finite step
strain $\gamma$, normalized by the relaxation $\sigma_{xy}(0,t)$ that
would occur in the linear limit of zero strain. \change{In strongly
  entangled systems anomalous behaviour in $h(\gamma,t)$ has been
  reported \citep{osaki1993damping,venerus2005critical}, corresponding
  to faster relaxation than predicted by DE theory. As noted above,
  dramatic heterogeneities can also occur during relaxation after a
  step strain, in the DRP model \citep{adamsolmsted09a} and
  experiments on entangled polymers
  \citep{wang2006nonquies,WangMM2007,boukany2009step}. These
  heterogeneities will have a signature in the stress response to a
  step strain. Thus, it is conceivable that transient heterogeneities
  could account for the anomalous damping function measurements
  \cite{marrucci1983free}.}

Hence, one must take great care in making predictions for constitutive
models: perturbations such as noise or the stress gradients of curved
geometries, or the presence of non-monotonic constitutive behavior,
necessitates a validation against a full inhomogeneous
calculation. This warning was also given by \citet{cook08}. Such
calculations are currently out of reach for models such as the GLAMM
model, which already involve solving partial differential equations in
time and arc length coordinates \citep{graham2003microscopic}.

\section{Summary}

Experiments on highly entangled polymers dissolved in their oligomers
show inhomogeneous transient velocity profiles in shear flow in a
variety of experimental tests. We have modeled this behavior using the
single mode diffusive Rolie-Poly (DRP) model. The DRP model has some
quantitative shortcomings, but can explain many of the experimental
transient phenomena. By varying the model parameters, namely the
convective constraint release efficiency $\beta$, the entanglement
number, $Z$ and the solvent viscosity $\epsilon$, we have shown that
transient shear banding can occur even for monotonic constitutive
curves. Similar behavior was found by \citet{cook08} in calculations
using a different non-monotonic model for entangled polymers, the
partially extended convective strain model.

A linear stability analysis of the startup transient shows that the
homogeneous state is unstable, at early times, to the inhomogeneous
transient shear bands. The ``weight'' $\Omega_{\mathrm{max}}$ of this
instability, given by the time integral of the most unstable
eigenvalue, provides a good predictor of flow
instability. $\Omega_\mathrm{max}$ is largest where the constitutive
curve is flat. If this weight is large then small perturbations can
provide the seed for this instability. For large enough perturbations,
particularly in fluctuations of the normal stress, the flow profile
develops a shear banding state, which then decays over a few
relaxation times $\tau_d$ once the eigenvalue returns to a stable
value.  We have shown that spatial perturbations play a crucial role
in triggering the transient instability. Examples include strong
enough stress gradients, as found in rotational rheometers; or
inhomogeneous initial conditions due to thermal noise; or residual
stress from loading the sample.

Hence, shear banding ranges from the steady state shear banding seen
in non-monotonic constitutive curves, to transiently inhomogeneous
flow that can develop even for monotonic constitutive curves.  The
stress overshoot during startup relaxes more quickly, due to the
transient shear banding, than if the fluid were to remain
homogeneous. This may be consistent with the suggestion long ago by
\citet{marrucci1983free} that an elastic instability could lead to
inhomgeneities in a step strain experiment.  The relaxation of the
stress overshoot is more pronounced in geometries with stronger stress
gradients.

Our calculations suggest that fluids with a non-monotonic
instantaneous constitutive curve $\sigma_{xy}(\dot{\gamma}, t)$ are more
likely to have an instability and transient banding
\citep{hayes2010constitutive}. Indeed, the linear stability analysis in
such situations leads to a unstable eigenvalue that is coincident (in
time) with the more strongly non-monotonic instantaneous constitutive
curves. Similarly, transient banding is strongly correlated with the
stress overshoot observed during startup \citep{sui2007iep}. This is a
subject for future work.

In attempting to fit the parameters of the Rolie-Poly model to the
existing data, it is apparent that the model should be judged mainly
for its qualitative conclusions: for example, shear banding in the
model occurs for much larger $Z$ than is found in experiments. The
model also fails to describe the high shear rate branch correctly
without violating certain physical criteria used to define the effect
of stretch on convective constraint release in the parent GLAMM model
\citep{likhtmangraham03,graham2003microscopic}. Finally, challenges
for future work include explaining the dramatic fracture seen in some
recent experiments on both entangled melts and solutions
\citep{boukany2009step,WangMM2007}, \change{and extending (or
  replacing) the GLAMM model to accurately describe the highest shear
  rate behavior.}

\acknowledgments We are grateful to the UK EPSRC (SMF, EP/E5336X/2)
and the Royal Commission of 1851 (JMA) for financial support; and to
Ron Larson for useful comments.
\bibliography{mastershear,articles,books}
\end{document}